\documentclass[preprint]{aastex}
\def\simgr{\,\hbox{\hbox{$ > $}\kern -0.8em \lower 1.0ex\hbox{$\sim$}}\,}
\def\simle{\,\hbox{\hbox{$ < $}\kern -0.8em \lower 1.0ex\hbox{$\sim$}}\,}

\shortauthors{THORSTENSEN, FENTON, \& TAYLOR.}
\shorttitle{Longer-Period Cataclysmics}
\begin{document}
\title{Spectroscopy of Seven Cataclysmic Variables 
with Periods Above Five Hours
\footnote{Based on observations obtained at the MDM Observatory, operated by
Dartmouth College, Columbia University, Ohio State University, and
the University of Michigan.}
}

\author{John R. Thorstensen and William H. Fenton}
\affil{Department of Physics and Astronomy\\
6127 Wilder Laboratory, Dartmouth College\\
Hanover, NH 03755-3528;\\
john.thorstensen@dartmouth.edu}
\author{Cynthia J. Taylor}
\affil{The Lawrenceville School\\
P.O. Box 6008, Lawrenceville, NJ 08648
}

\begin{abstract}
We present spectroscopy of seven cataclysmic variable stars with
orbital periods $P_{\rm orb}$ greater than 5 hours,
all but one of which are known to be dwarf novae.
Using radial velocity measurements we improve on previous 
orbital period determinations, or derive periods for the first time.  
The stars and their periods are
TT Crt, 0.2683522(5) d; 
EZ Del, 0.2234(5) d; 
LL Lyr, 0.249069(4) d;
UY Pup, 0.479269(7) d; 
RY Ser, 0.3009(4) d; 
CH UMa, 0.3431843(6) d; 
and SDSS J081321+452809, 0.2890(4) d.
For each of the systems we detect the spectrum of the 
secondary star, estimate its spectral
type, and derive a distance based on
the surface brightness and Roche lobe constraints.
In five systems we also measure the radial
velocity curve of the secondary star, estimate 
orbital inclinations, and where possible
estimate distances based 
on the $M_V({\rm max})$-$P_{\rm orb}$
relation found by Warner.  In concordance with
previous studies, we find that all the secondary stars
have, to varying degrees, cooler spectral types 
than would be expected if they were on the main sequence at the
measured orbital period.
\end{abstract}

\section{Introduction}

Cataclysmic variable stars (CVs) are close binaries in which a white
dwarf accretes matter from a less evolved companion (the secondary),
which usually resembles a lower-main-sequence star.  CVs have a rich
phenomenology, and the theory of CVs involves contributions from many
subdisciplines of astrophysics. \citet{warn} presents a comprehensive
review. 

Mass transfer in CVs occurs through Roche lobe overflow, so the 
secondary's surface coincides almost exactly with the Roche lobe.
This constrains the secondary's mean  
density to be nearly fixed at a given orbital period
$P_{\rm orb}$.  Because the mean density is a strong function of mass
along the lower main sequence, there is a correlation
between the secondary's spectral type and the orbital period, with
hotter, more conspicuous secondaries appearing at longer $P_{\rm orb}$. 

In many instances this simple picture is not accurate.  As a cataclysmic
evolves through mass transfer to shorter $P_{\rm orb}$, the secondary
star's thermal timescale grows comparable to the orbital evolution time,
leading to departures from thermal equilibrium and consequent
discrepancies between reality and expectations based on main-sequence
characteristics.  Even so, \citet{beuermann2ndry} found that the
relatively small number of CV secondaries detectable in systems with
$P_{\rm orb} < 3$ h are fairly close to the main sequence, so these
departures are evidently not severe in these cases.  In contrast, a
substantial fraction of {\it longer}-period secondaries are
significantly cooler than expected; at any given $P_{\rm orb}$, the
hottest secondaries lie near the main sequence, but there are many
cooler objects.  This suggests that in many longer-period CVs, the
secondaries have begun nuclear evolution prior to mass transfer
\citep{bk00}.  Further support for this scenario comes from
\citet{gaensicke03}, who point out anomalous
nitrogen-to-carbon ratios in several CVs, indicating that the material
being transferred has passed through CNO burning.  Also,
the anomalous objects EI Psc (= RX
J2329+06) and QZ Ser have quite short periods (64 min and 2.0
hr respectively), yet show strong K-star features in their spectra; these
secondaries are much too {\it early} for their orbital period.  QZ Ser in
addition shows enhanced sodium features which may indicate CNO-processed
material \citep{thoreipsc,thorqzser}.  These anomalous systems are
nicely matched by models in which mass transfer begins at longer periods
after the onset of nuclear evolution.  Their antecedents should
be anomalously cool systems at longer periods.

Studies of longer-period systems can therefore provide useful empirical
clues to CV evolution.  Furthermore, when the secondary star's
contribution is visible, as it often is at longer periods, it offers
practical advantages and opens several lines of investigation, as
follows: (1) Because the secondary's orbit is accurately circular, and
its surface is free of significant non-orbital motion, the radial
velocities of the secondary tend to be much better-behaved than emission
line velocities, which can arise from complicated, variable flows of
gas.  It is easier to find $P_{\rm orb}$ from an accurately sinusoidal 
velocity curve than from the jittery velocities often found for 
emission lines.  (2)
Similarly, the secondary's velocity curve should track its
center-of-mass motion in a relatively accurate and straightforward
manner.  Complications can arise from nonuniformity of the secondary's
surface (especially asymmetries between the secondary's front and back
sides), but these are less severe than the gross difficulties often
presented by emission lines. (3) If the emission-line velocities are to
trace the white dwarf's motion, it is necessary (but not sufficient) for
their phase to be one-half cycle away from the secondary's motion.  In
non-eclipsing CVs there is often no absolute marker of binary phase, but
when secondary star radial velocities are measurable, they provide just
such a marker.  If the emission and absorption phases differ by 180
degrees, as expected, it is at least {\it possible} that the emission
lines follow the white dwarf motion; if so, a mass ratio can be computed.  (4)
The kinematics of the CV population provide a clue to their age and
evolutionary status \citep{kolbstehle}.  As \citet{north02} point out,
the secondary spectra can provide systemic radial velocities which are
more reliable than those inferred from emission lines, provided the
zero point is established carefully.  (5) At a given orbital period, the
Roche lobe constraint makes the secondary's radius a weak function of
its mass.  An accurate spectral type for the secondary constrains its
surface brightness, which together with the radius yields an estimate of
the absolute magnitude.  If the secondary's contribution to the system's
light can be measured, this gives a distance estimate. 

In this paper we present studies of seven longer-period systems.  In all
cases the $P_{\rm orb}$ is either determined for the first time or
improved.  We detect the secondary stars and, insofar as possible,
perform the analyses outlined above.  Section 2 outlines the
methodology, Section 3 details results for the individual stars, and
Section 4 is a brief discussion.

\section{Techniques}

Table 1 is a journal of our observations.  We obtained spectra at the
MDM Observatory at Kitt Peak, Arizona, mostly with the 2.4m Hiltner
telescope, with a few observations from the 1.3m McGraw-Hill telescope.
At the 2.4m we used the `modular' spectrograph, a 600 line mm$^{-1}$
grating, and a SITe 2048$^2$ CCD detector yielding 2 \AA\ pixel$^{-1}$
from 4210 to 7560 \AA\ (vignetting severely toward the ends), and
typical resolution of 3.5 \AA\ FWHM; at the 1.3m, we used the Mark III
spectrograph and a $1024^2$ SITe CCD giving 2.2 \AA\ pixel$^{-1}$ from
4480 to 6780 \AA, with 4 \AA\ FWHM resolution.  A 1-arcsec slit was used
at the 2.4 m, and a 2-arcsec slit at the 1.3 m.  In order to maintain an
accurate wavelength calibration we took exposures of comparison lamps
hourly as the telescope tracked, and whenever the telescope was moved.
The measured wavelength of the $\lambda 5577$ night-sky line provided a
check; it was generally stable to less than 10 km s$^{-1}$ in the
reduced spectra.  For a few of the 1.3m spectra we were unable to take
comparison lamps because of equipment trouble, and we calibrated these
by using the night sky lines to find a zero-point offset for each
spectrum.  

When the sky was suitably clear, we took spectra of flux standards to
derive the instrument response function.  We also observed bright B
stars and used these spectra to derive an approximate correction for
telluric absorption bands.  For most of our 2.4 m observations, we
rotated the spectrograph slit to the parallactic angle to avoid
dispersion losses.  Even so, our individual 2.4 m data often showed
unphysical, irreproducible fluctuations in the continuum shape, which we
still do not understand; however, these appear to average out over many
exposures.  Such averaging occurs both in the computation of the
instrument response function and the compilation of our mean
program-object spectra, so the continua in our mean spectra should be
reasonably accurate.

For our reductions we used standard IRAF\footnote{IRAF is distributed by the
National Optical Astronomy Observatories.} procedures for flatfielding,
extraction of one-dimensional spectra, wavelength calibration, and flux
calibration.

Our main goal was to determine orbital periods for these systems from
radial velocities.  To measure emission line velocities (almost exclusively
H$\alpha$) we used tunable convolution algorithms described by
\citet{sy80}.  We also computed a velocity error estimate by propagating
the counting-statistics errors generated by the IRAF reductions.  In
most cases the scatter of the data around the best-fitting sinusoid
significantly exceeded the counting-statistics estimates, indicating
that the emission line profiles were not particularly well-behaved.  

In some cases we tried to extract dynamical information from the
emission lines using the technique pioneered by \citet{shafter83}
(hereafter the `Shafter diagram' technique).   In
this, one convolves the emission line with an antisymmetric function
consisting of a positive and a negative Gaussian with a tunable
separation $\alpha$, and repeats the measurement and sinusoidal fit with
increasing separations until the velocity errors become excessive.  The
sinusoid fitted to the velocities measured with the widest separation at
which the errors are reasonable is taken to represent the motion of the
white dwarf.  This procedure measures the high-velocity wings of the
line, which are produced close to the white dwarf, where one hopes that
the rapid rotation of the disk smooths out asymmetries.  There are many
cataclysmics in which this hope is clearly not realized; when an
absolute phase marker is available (such as an eclipse), the phase of
the emission-line velocities in some systems does not track the known
phase of the white dwarf's motion.  In most of the stars studied here we
do have an absolute phase marker -- the secondary star's velocity curve
(see below) -- so the emission-line phases can be checked to see if they
at least agree with expectation.  Such agreement is a necessary, but not
sufficient, condition for the emission velocities to be dynamically
faithful.  We use them, but caution the reader that they may not be
quite right.

None of the systems studied here are known to eclipse, so their
inclinations are not known with enough precision to find 
accurate masses.  However, as noted earlier the absorption-line
radial velocities give a relatively straightforward measure of the
secondary's projected  orbital velocity, and this can be used to establish
constraints on the orbital inclination, provided the masses are within
realistic limits. \citet{warn87} found a relationship in dwarf novae
between $P_{\rm orb}$ and the inclination-corrected absolute magnitude at maximum light, 
$M_V ({\rm max})$. In several cases here
our dynamically-constrained inclinations are accurate enough 
to use this relationship to estimate distances. 

K-type secondaries were evident in several of these objects.  For these
we measured absorption-line velocities using the {\it xcsao}
cross-correlation radial velocity package \citep{kurtzmink}.  We mostly
correlated the range from 5050 to 6500 \AA , excluding the complicated
region around NaD (which usually included HeI $\lambda5876$ emission);
for some objects restricted the correlation to the 6000 to 6500 \AA\ region.  
The {\it xcsao} task produces an error estimate based on the $R$-statistic developed by
\citet{tonrydavis}; in well-exposed spectra the estimated uncertainty was often 
$\simle$ 10 km s$^{-1}$.   The scatter of the velocities around the best sinusoidal
fits was generally comparable to the Tonry-Davis errors, indicating that
they are fairly realistic.  

For the cross-correlation template spectrum, we used a
velocity-compensated sum of 86 observations of IAU velocity standards,
accumulated over the past several years with the same equipment and
procedures used for the program objects.  The standards were all K or
late G stars, mostly giants.  When the individual velocity standard
spectra were cross-correlated against this sum, the RMS scatter of the
observations around the cataloged velocities was 7.1 km s$^{-1}$.  
The photon statistics of the standard star observations were always
excellent, so this test is dominated by systematic effects, including
(1) spectral mismatch of the individual
stars compared to the sum, (2) imperfections in the wavelength
calibration, and (3) mis-centering of the star in the spectrograph slit,
leading to spurious wavelength shifts.  
Because of the large number of standard
observations, we estimate that the zero point is determined to $\sim 2$
km s$^{-1}$.   The stars below typically have dozens of observations, so
the formal errors on their $\gamma$ (i.e., mean systemic) velocities are
often very small, but we estimate the external accuracies of $\gamma$ to
be $\sim 5$ km s$^{-1}$; spectral mismatches, subtle asymmetries in the
surfaces of the CV secondaries, and other imponderables probably enter
around this level.

For period searches, we used a `residual-gram' technique described in
\citet{tpst}.  This works especially well for data which follow a
sinusoid accurately.  Because the time sampling was nonuniform, numerous
alias periods typically turn up in the period searches, corresponding to
differences in the cycle count during gaps in the data.  These manifest
as candidate frequencies separated by $1/T$, where $T$ is the length of
the gap; $T = 1$ d, for example, corresponds to the daily cycle count
ambiguity.  In some of the objects the choice of alias was clearly
unambiguous, but where needed we applied the Monte Carlo procedure described
by \citet{tf85} to determine the confidence with which the best-fitting
period could be identified with $P_{\rm orb}$.  Once a period was
adopted, the variation was fitted with sinusoids $$v(t) = \gamma + K
\sin[2 \pi(t - T_0)/P]$$ using a hybrid linear least-squares algorithm.
Note that $T_0$ is the epoch of apparent inferior conjunction of  the
source being observed.  If the source is the secondary, $T_0$ is the
epoch at which eclipses of the white dwarf would be expected if 
the system were edge-on.  

To characterize the secondary's spectral contribution, we began by
shifting the individual spectra into the rest frame of the secondary
spectrum, using the fitted sinusoidal orbit and the {\it rvsao} task
{\it sumspec}, and then averaged these.  We have a set of spectra of
K-dwarfs classified by \citet{keenan89}, observed with our standard 2.4
m instrumentation, and a similar set of M-dwarf spectra
classified by \citet{boe76}.  We shifted these to zero velocity, scaled each
spectrum using a range of multiplicative factors, and subtracted the
scaled spectra from the program object's velocity-compensated spectrum.
Finally, we examined all these subtracted spectra by eye and decided on
a range of acceptable spectral types and flux contributions for the
secondary, based on how well the secondary's spectral features had been
removed from the original spectrum.  

Tables in \citet{beuermann99} can be used to find surface brightnesses
as a function of spectral type, and expressions in
\citet{beuermann2ndry} yield an estimated size of the secondary star
given the orbital period and an estimate of the secondary's mass.  The
secondary mass is generally unknown, but the radius depends only on its
cube root; as a guide, we use the evolutionary calculations of
\citet{bk00} to estimate a range of plausible secondary masses at the
system's $P_{\rm orb}$ and secondary spectral type.  The surface 
brightness, radius, and flux then yield a distance estimate 
Note that 
this procedure does {\it not} assume that the absolute magnitude of 
the secondary is appropriate to its spectral type
on the main sequence, but only that the secondary's {\it surface
brightness} is appropriate to its spectral type.

To compute the magnitude of the secondary star alone, we applied the
IRAF {\it sbands} task to the subtracted K-star spectrum.  This task
synthesizes magnitudes from a flux-calibrated spectrum using a passband,
in this case the $V$ passband tabulated by \citet{bessell}.  To check
this procedure we synthesized $V$ magnitudes for the flux standard stars
and the bright B stars used to map the continuum and band shapes.  Over
several observing runs the rms variation of the synthetic-minus-catalog
magnitudes for these stars was 0.15 mag, with mean zero-point offsets of
less than 0.1 mag.  These stars may have been centered in the slit a
little more attentively for these brief exposures than for the longer
sequences on the program stars, and conditions may have been better on
average 
so we adopt 0.25 mag as the contribution of the calibration to
the error budget.  Because standard star observations were only taken
when the sky appeared clear, but program star observations were taken
whenever possible, we assume that our secondary 
stars are $\sim 0.2$ mag brighter than our calibrations would 
indicate. 

Several of these systems are substantially reddened.  For all of the 
distance calculations, we used the \citet{schlegel98} maps to estimate
the reddening, sometimes reducing the implied $A_V$ if it appeared
likely that the star did not lie entirely outside the Galaxy's dust layer.
In most cases the uncertainty in the reddening contributed 
little to the error budget.

%When adequate phase coverage was available, we explored the behavior of
%the spectrum around the orbit by rectifying our spectra and assembling
%them into two-dimensional greyscale phase-resolved images, using a
%procedure similar to that outlined by \citet{taylor99}.  This can be 
%especially revealing for emission-line behavior.

\section{The Individual Stars}

The results for all the stars are summarized in tables and figures.
Table 2 gives parameters of the emission lines measured from the mean
fluxed spectra; in many cases the NaD absorption lines, blended at this
resolution, are included as well.  Table 3 
lists all the radial velocities.  Table
4 gives parameters of sinusoidal fits to the velocities, and Table 5
gives derived characteristics of the secondary stars and summarizes a
secondary-based distance estimate for each system.  Fig.~1 shows the
mean fluxed spectra of all the objects, and Fig.~2 shows the 
folded radial velocity curves.

\subsection{TT Crt}

\citet{szkodyttcrt} obtained spectroscopy and photometry of this
system and found candidate orbital periods of 438 and 445 min.  They
detected the secondary star in the spectrum, and estimated a spectral
type of K5 - M0.  Time-series photometry showed apparent ellipsoidal
variations.  

Our observations of TT Crt span 1262 d.  The absorption velocities
follow a sinusoid with large amplitude and little scatter, and the 
cycle count over the whole interval is determined without ambiguity, 
yielding $P_{\rm orb} = 0.268351(1)$ d, or 386 min.  The H$\alpha$ 
emission velocities,
while less accurate, independently constrain the period to the same
value within the uncertainties.  Although the candidate periods found by
\citet{szkodyttcrt} are similar to the more reliable period
found here, they are formally inconsistent with the present result
and appear to reflect a mistaken cycle count over a 2-day interval.
\citet{szkodyttcrt} do give an epoch of apparent inferior conjunction of
the secondary star in 1991 February, and our period is precise enough
to extrapolate back to their observation without ambiguity, yielding
a refined ephemeris, 
$$\hbox{Secondary inferior
conjunction} = \hbox{HJD 2,452,297.025(1) + 0.2683522(5)} E,$$ 
where
$E$ is an integer.  

The flux-decomposition procedure yielded K5$ \pm 1$, toward
the early end of the range found by \citet{szkodyttcrt}.  
As \citet{szkodyttcrt} pointed out, the emission lines in TT Crt
are double-peaked most of the time; H-alpha shows a separation 
of $\sim 670$ km s$^{-1}$.  

TT Crt has the largest absorption-line velocity amplitude 
$K_{\rm abs}$ of the systems studied here.  
The Shafter diagram showed the emission lines to be accurately
antiphased to the absorption, and to have a velocity
amplitude $K_{\rm emn}$ fairly insensitive to the convolution
function width $\alpha$; we adopted $\alpha = 1800$
km s$^{-1}$ for the emission-line measurements, which yields
a nominal $q = M_2/M_1 = 0.58$.  Curiously, the fits show
the emission line mean velocity $\gamma_{\rm em}$ to be a function of 
$\alpha$, and $\gamma_{\rm em}$ disagrees significantly with the corresponding
value for the absorption lines $\gamma_{\rm abs}$.  For the reasons noted earlier
we believe $\gamma_{\rm abs}$ to be the more reliable measure
of the systemic velocity.  
The photometry presented by
\citet{szkodyttcrt} is extensive enough to rule out any
significant eclipse, which constrains the inclination $i$ to be
less than about 70 degrees to avoid an obvious partial eclipse
of the disk.  At $i = 70$ degrees the minimum white dwarf
mass is around 0.8 M$_{\odot}$.  A comfortable fit to all the
data occurs for $M_1 = 1.0$ M$_{\odot}$ and $i = 60$ degrees.
The inferred white dwarf mass exceeds the Chandrasekhar limit
for $i \simle 52$ degrees.   As always, these constraints could 
be relaxed slightly if the measured $K_2$ misrepresents the 
secondary's center-of-mass motion, but the inclination is 
rather tightly constrained, and it is unlikely that the
$M_1$ is less than 0.7 M$_{\odot}$.  Also, the estimate of the 
secondary's mass used to estimate the Roche lobe size in the 
distance calculation,
$M_2$ = 0.5 - 0.8 M$_{\odot}$,
is consistent with the mass ratio only if the white dwarf is
$\sim 1$ M$_{\odot}$.   While there is no guarantee that 
our adopted secondary mass is 
correct, the distance estimate fortunately depends
only weakly on the assumed $M_2$.  

At $i=60 \pm 10$ degrees, the \citet{warn87} $M_V$-$P_{\rm orb}$ 
relation predicts $M_V = 4.1 \pm 0.5$, where the uncertainty
shown is purely from the uncertainty in the inclination.  
The General Catalog of Variable Stars (GCVS; \citealt{gcvs}) 
gives $V_{\rm max} = 12.5$, and the extinction is 
about 0.1 mag, yielding $(m-M)_0 = 8.3$ mag, or 460 pc.
The estimate based on the secondary is $9.4$ mag, or 
760 pc.  These are broadly in agreement, but a little more
discrepant than might be expected. 

%The periods cited by \citet{szkodyttcrt} corresponds to 
%$\sim 3.27$ cycle d$^{-1}$, while the present period is 
%$\sim 3.73$ cycle d$^{-1}$, so the previous determination
%may have suffered from a cycle count error over a two-day
%baseline.

% \citet{akaavso}

% UCAC: 11 34 47.172  -11 45 30.29, significant PM mura   -24.5 mudec    31.2 mas/yr
% My own stuff (w/ Diana): PM star  1 :  11 34 47.219   -11 45 30.01  
% mura  -21 +-    4 mudec =   25 +-    4

\subsection{EZ Del}

EZ Del was misidentified in the \citet{ds92} atlas. \citet{liupap1}
found that a star 7 arcsec SE of the marked star showed a
typical CV spectrum, and the correct star is marked in \citet{livinged}.

The 2002 June velocities are best fit by a frequency near 5.5 cycle
d$^{-1}$, while the 2003 June velocities indicate 4.5 cycle d$^{-1}$.
The 2003 June observations were timed to improve discrimination of
the cycle-count aliases, and the 4.5 cycle d$^{-1}$ gives the
better fit to the combined data, which has a discriminatory power
(defined by \citealt{tf85}) near 0.98.  The daily cycle count is
therefore fairly clear but not absolutely secure.  The period listed
in Table 4 is the weighted average of periods found in separate fits 
to the two 
observing runs; the (unlikely) alternate daily cycle count would give
$P_{\rm orb} = 0.1822(5)$ d.  An unknown of cycles elapsed in 1-year
gap between observations, leaving the precise period ambiguous.
Allowed periods lying within $\pm 4$ standard deviations of the period
in Table 4 are given by $P = 373.830(3) {\rm d} / N$, where the
integer $N = 1673 \pm 15$. 

The mean fluxed spectrum of EZ Del is quite blue, and the mean spectra
from 2002 June and 2003 June differ significantly.  The 2002 June
spectrum had a synthetic $V$ magnitude of 17.6, and a 4500 to 7500
\AA\ continuum fit well by $f_{\lambda} \propto \lambda^{-1.76}$,
while the 2003 June spectrum had 18.0 and a power-law exponent of
$-0.92$, still rather bluer than most dwarf nova at minimum light.  It
thus appears that neither of our spectra were taken fully at minimum.
Exploratory spectra taken 1999 June 10.4 UT and 2001 June 24.3
UT both showed the star in outburst, with synthetic $V$ magnitudes of
15.3 and 15.7 respectively; these limited observations suggest that EZ
Del outbursts frequently.

A late-type secondary star is detected in the spectrum, but because of
the modest signal-to-noise and the secondary's small fractional
contribution, it is subtle; only the broad TiO bands are visible at
low amplitude.  We could not measure radial velocities of the secondary
star, and with no absolute phase marker we choose not to infer dynamical 
information from the emission line velocities. 

\subsection{LL Lyr}

\citet{smith2ndary} detected the TiO bands and Na $\lambda 8190$ doublet
of the secondary star.  Our summed, fluxed spectrum shows clear features
of an M-dwarf secondary and relatively broad emission lines whose peaks
are not quite doubled.  
The system's synthetic $V$ magnitude is 17.8.  

The emission-line velocities of LL Lyr are from six observing runs
spanning 1083 d, and define an unambiguous cycle count over the entire
interval, yielding $P_{\rm orb} = 0.249069(5)$ d.  While the 
signal-to-noise ratio of the secondary star's spectrum was inadequate
for velocity measurements, we could again estimate the secondary
contribution and infer a distance.   Our spectral type estimate,
M2.5 $\pm$ 1.5, is in satisfactory agreement with the M3 - M4
classification found by \citet{smith2ndary}.   LL Lyr is an example
of a long-period dwarf nova with a secondary which is considerably
later than would be expected for the main sequence.

Again, without velocities of the secondary, we do not attempt a 
dynamical analysis.

\subsection{UY Pup}

\citet{lockleyuy} obtained spectra in outburst and estimated 
$P_{\rm orb} = 10.22 \pm 0.19$ h. They inferred a low orbital
inclination.

UY Pup outbursts frequently, and the mean spectrum shows a
blue continuum and a relatively weak secondary contribution
despite the long orbital period.  Our mean spectrum
evidently was taken when the system was above minimum light.

The time series spans 603 days, and we find consistent, 
unambiguous periods in both the emission and absorption lines,
the weighted mean being 0.479269(7) d, or 11.50 h.  Thus the 
previous period is again similar to, but formally inconsistent
with, the more accurate period found here.
The secondary star, for which we find K2 - K6, is clearly
not on the main sequence, since a main sequence secondary in 
this period regime would be much earlier \citep{lockleyuy}.

Because $K_2$ is fairly small at $102 \pm 4$ km s$^{-1}$, the
inclination is low and is confined to a rather narrow range.  The
Shafter diagram shows $K_{\rm em}$ increasing with the convolution width
$\alpha$.  If we adopt $\alpha = 1200$ km s$^{-1}$, then $K_{\rm em}$ =
96 km s$^{-1}$, and the mass ratio $q = M_2 / M_1$ is fairly close to
unity, but the sensitivity of $K$ to $\alpha$ suggests that this is not
particularly reliable.  A 0.8 M$_{\odot}$ white dwarf and 0.6 M$_{\odot}$ secondary
at $i \sim 40$ degrees fits all the data nicely.  If we demand that $q <
1$, and that $M_1 > 0.5$ solar masses, then $i < 50$ degrees; keeping
the white dwarf mass well below the Chandrasekhar limit requires $i >
25$ degrees.  

Adopting $i = 38$ degrees, the \citet{warn87} relation yields
$M_V({\rm max}) = 2.3$, and because
the inclination correction flattens out toward low
inclination the resulting uncertainties
are small.  The GCVS gives the photographic magnitude
at maximum as 13.0.  Since dwarf novae in outburst have
small color indices, we take this as $V_{\rm max}$ and 
assume $A_V = 0.5$ to find $(m - M)_0 = 10.2$, or
$d = 1100$ pc.  This agrees very well with the estimate
from the secondary star (Table 5).

%\citet{akaavso}
%\citet{lockleyuy}

% UCAC2 Coords:   7 46 31.241  -12 57 08.99 r =  13.04 b =   0.00
% mura    -7.8 mudec    11.1

\subsection{RY Ser}

This dwarf nova was mis-identified in the \citet{dws97} atlas, but is correct
in the on-line {\it Living Edition} of the atlas \citep{livinged}.
Perhaps because of its historically uncertain identification, relatively 
little information is published.

The mean flux-calibrated spectrum implies $V \sim 16.4$, relatively bright for
an unstudied dwarf nova.  The secondary star is clearly detected.  The emission
lines are relatively narrow and single-peaked.  

Although our observations of this system are sparse (only 14
exposures in a single run), both the secondary star's velocities and the
H$\alpha$ emission velocities unambiguously indicate a period near 0.30 days,
the weighted mean being 0.3009(4) d.   Because $K_{\rm abs}$ is 
small at $87 \pm 6$ km s$^{-1}$, the inclination must be quite low.  From 
the Shafter diagram we adopted $\alpha = 1060$ km s$^{-1}$, yielding 
$K_{\rm em} = 73 \pm 10$ km s$^{-1}$, implying $q \sim 0.8$.  
Demanding $M_1 > 0.5$ constrains $i$ to be less than about 31 degrees.
A good fit is obtained at $M_1 = 0.8$ M$_{\odot}$, $M_2 = 0.6$ M$_{\odot}$, and
$i = 26$ degrees.

The \citet{warn87} relation predicts $M_V(\rm max) = +3.1$ with little 
uncertainty introduced by the inclination.  For $V_{\rm max} = 13.1$ 
\citep{aavso},
and a very substantial $A_V = 1.2$, this gives $(m - M)_0 = 8.8$, 
or 580 pc.  The secondary star gives a distance in close agreement.

% object mis-IDd in DWS97, but correct in live version.  It's listed in 
% UCAC-2 at 17 23 07.176  -12 48 09.78.  PM is (-4,-2) mas/yr, 
% i.e., basically zero.

\subsection{CH UMa} 

\citet{thorpg} found an 8.3-hour radial velocity period in the Balmer lines in
CH UMa, but could not exclude a 12.5-hour daily alias.  \citet{friendchuma}
obtained radial velocities of the secondary by cross-correlation near the 8190
\AA\ Na doublet, and confirmed the 8.3-hour alias choice.  They also found a
statistically significant orbital eccentricity $e = 0.10$, which is unexpected
in a CV, the orbits of which are expected to circularize quickly.

The mean fluxed spectrum shows a strong secondary contribution and quite
narrow emission lines.  The synthetic magnitude is $V = 15.2$. 

Because of the brightness and the strong secondary
contribution, the 28 cross-correlation velocities of CH UMa are
relatively precise, with mean Tonry-Davis errors near 10 km s$^{-1}$ and
an rms scatter about the best sinusoidal fit near 8 km s$^{-1}$.  Our
data span 1259 d and, taken alone, constrain the period to 0.343181(4) d, with no
ambiguity in cycle count.  \citet{friendchuma} give an epoch of
red-to-blue crossing of the absorption velocity, which can be connected
to our fits without ambiguity, yielding the ephemeris 
$$\hbox{Secondary inferior conjunction} = \hbox{HJD 2,452,442.788(3) + 
0.3431843(6)} E,$$
where $E$ is an integer.

We see no indication of the eccentricity in the secondary-star orbit
noted by \citet{friendchuma}.  The residuals as a function of phase
showed no trends suggesting eccentricity.  Fitting the velocities with
an eccentric orbit yielded a best-fit $e = 0.012$, insignificantly
different from zero, and the fit improved only marginally.  Fixing
the eccentricity held at their best-fit $e = 0.10$ led to slightly
worse fits than with a circular orbits.  We have
somewhat fewer measurements than \citet{friendchuma} (28 to their 38),
but our accuracy appears to be comparable, so the non-confirmation
has some weight; we cannot rule out a nonzero eccentricity but we
regard it as somewhat unlikely.  A definitive test
would require more phase coverage and better radial-velocity accuracy.
Our fitted velocity amplitude $K = 76 \pm 3$ agrees well with that
measured by \citet{friendchuma} ($78 \pm 3$), but our mean
velocity $\gamma$ disagrees slightly ($-15 \pm 2$ against their
$-3 \pm 3$).

% 2000A&A...354..103S - Astron. Astrophys., 354, 103-111 (2000) - February(I) 2000
% Outburst parameters and the long-term activity of the dwarf nova CH Ursae Majoris.

\subsection{SDSS J081321+452809}

This object (hereafter SDSS0813) is one of the substantial number 
of CVs turned up by the Sloan Digital Sky Survey.  \citet{szkodysdss1} 
presented a spectrum, noted the presence of a late-type companion, 
and suggested that the orbital period was likely to be fairly long.  

Our mean fluxed spectrum closely resembles that found by 
\citet{szkodysdss1}, with relatively narrow emission lines and 
a strong secondary spectrum.  The synthetic $V = 18.4$ is similar to the 
$g^* = 18.29$ measured in the Sloan survey \citep{szkodysdss1}.

We have 26 spectra; H$\alpha$ emission was measurable
in all of them, and usable absorption velocities were found
for all but three.  A single spectrum was obtained 2002 Feb.\ 19, and
the remainder were obtained on another observing run 30 days
earlier.  When the Feb.~19 point is omitted, the remaining 
emission and absorption velocities give similar periods, with
no daily cycle count ambiguity; their weighted average is 0.289(1) d.
The Feb.~19 point introduces cycle count ambiguity across
the 30-day gap, the most likely choice yielding
0.2890(4) with 0.2867(4) d being somewhat less favored.  
Szkody et al.'s suggestion of a long period is evidently correct.

The distance based on the secondary star is around 2.1 kpc, which
at the rather high Galactic latitude (32.9 deg) puts the system
over 1 kpc from the Galactic plane.  The distance uncertainty quoted in
Table 5 is a quadrature sum of the various uncertainties.  To establish
a smallest plausible distance, we skew all the contributing quantities
in the sense minimizing the distance, and find 1500 pc, which still puts
the system 800 pc from the plane.  

This system is notable as a long orbital period CV which is not
known to erupt.  If it were to undergo dwarf nova eruptions, 
the \citet{warn87} relation predicts it would reach $V = 14.7$ at
its inclination and distance.  It is conceivable that 
eruptions of this magnitude could have been overlooked, so 
a dwarf-nova classification remains a possibility.  

\section{Discussion}

As noted earlier, the secondaries in long-period cataclysmics are
mostly later than expected on the basis of main-sequence models,
with the edge of the observed distribution coinciding approximately with  
main-sequence expectations (\citealt{beuermann2ndry} and \citealt{bk00}
plot known systems on the spectral-type vs. $P_{\rm orb}$ diagram).
The present small sample is nicely consistent with previous results.
TT Crt comes the closest to the main-sequence
spectral type expected for its period, missing it by only two
subclasses, while UY Pup, CH UMa, and LL Lyr are all far cooler than
predicted by main-sequence models.  UY Pup is also notable for its
unusually long period.  

Because we detect and classify the secondary stars in all these 
systems, we can find distances based on the secondaries' surface
brightnesses. 
For the four dwarf novae in which the secondary velocity curve allows
us to also estimate the inclination, we derive an alternate
distance estimate using the \citet{warn87} $M_V({\rm max})$-$P_{\rm
orb}$ relation.  Reasonably good agreement is found.

We confirm the conjecture by \citet{szkodysdss1} that SDSS0813 is 
a long-period system; the inferred distance is therefore large,
and SDSS0813 lies far from the Galactic plane.  
Dwarf nova eruptions, if present, have been overlooked.
\citet{tappert01} discuss another long-period system
at high latitude, CW 1045+525, which is also not known to 
outburst.  Like SDSS0813, CW 1045+525 was not discovered through
its variability, but rather in the Case objective-prism
survey.  The SDSS, which selects peculiar objects through accurate
color photometry and follows them up with spectroscopy, may 
finally provide the basis for an accurate accounting of the 
CV population. 

{\it Acknowledgments.} 
The NSF supported this work through grants
AST 9987334 and AST 0307413.  We thank the MDM staff for
supporting the many observing runs on which these data were taken.

\clearpage
\tabletypesize{\footnotesize}
\begin{deluxetable}{lrcc}
\tablewidth{0pt}
\tablecolumns{4}
 
\tablecaption{Journal of Observations}
\tablehead{
\colhead{Date} &
\colhead{$N$} &
\colhead{HA (start)}  &
\colhead{HA (end)} \\
\colhead{[UT]}  &
 &
\colhead{[hh:mm]} &
\colhead{[hh:mm]} \\
}
\startdata

\cutinhead{TT Crt:} 
2000 Jan 6 &  2 & $ +1:26$ & $ +1:32$ \\ 
2000 Apr 7 &  1 & $ +1:53$ & \nodata \\ 
2001 Dec 18 &  2 & $ +0:04$ & $ +0:14$ \\ 
2001 Dec 25 &  3 & $ -2:49$ & $ -2:28$ \\ 
2001 Dec 27 &  2 & $ -2:27$ & $ -2:17$ \\ 
2002 Jan 19 &  3 & $ -1:37$ & $ +2:25$ \\ 
2002 Jan 20 &  2 & $ -0:20$ & $ +2:11$ \\ 
2002 Jan 21 &  4 & $ -3:32$ & $ +1:48$ \\ 
2002 Jan 22 &  4 & $ -3:01$ & $ +2:11$ \\ 
2002 Jan 24 &  3 & $ -1:27$ & $ +1:23$ \\ 
2002 Feb 16 &  2 & $ -0:18$ & $ -0:07$ \\ 
2002 Feb 20 &  1 & $ +3:39$ & \nodata \\ 
2002 Jun 12 &  1 & $ +1:58$ & \nodata \\ 
2002 Jun 13 &  1 & $ +2:13$ & \nodata \\ 
2002 Jun 14 &  1 & $ +2:23$ & \nodata \\ 
2002 Dec 13 &  1 & $ -0:13$ & \nodata \\ 
2003 Jun 20 &  1 & $ +2:41$ & \nodata \\ 
2003 Jun 21 &  1 & $ +2:44$ & \nodata \\ 
2003 Jun 22 &  1 & $ +2:32$ & \nodata \\ 
\cutinhead{EZ Del:}
2002 Jun 14 & 13 & $ -2:26$ & $ -0:29$ \\ 
2002 Jun 15 &  2 & $ -2:11$ & $ -2:02$ \\ 
2002 Jun 16 &  5 & $ -4:02$ & $ +0:50$ \\ 
2002 Jun 17 &  6 & $ -3:21$ & $ +0:32$ \\ 
2003 Jun 23 &  6 & $ -4:56$ & $ +1:17$ \\ 
2003 Jun 24 &  7 & $ -4:27$ & $ -3:31$ \\ 
2003 Jun 25 &  9 & $ +0:33$ & $ +1:38$ \\ 
\cutinhead{LL Lyr:}
2000 Jul 5 &  3 & $ +2:38$ & $ +4:06$ \\ 
2000 Jul 6 &  9 & $ -2:34$ & $ -0:59$ \\ 
2000 Jul 7 &  2 & $ -3:05$ & $ -2:55$ \\ 
2001 May 12 &  6 & $ -0:05$ & $ +0:17$ \\ 
2001 May 14 & 18 & $ -1:36$ & $ -0:35$ \\ 
2001 May 15 & 20 & $ -2:55$ & $ +1:03$ \\ 
2001 May 16 & 12 & $ -4:59$ & $ +0:47$ \\ 
2001 May 18 &  8 & $ -4:33$ & $ -3:25$ \\ 
2001 Jun 26 &  1 & $ +1:22$ & \nodata \\ 
2001 Jun 27 &  6 & $ -3:42$ & $ +3:23$ \\ 
2001 Jun 28 &  3 & $ -3:48$ & $ -3:30$ \\ 
2001 Jun 29 &  3 & $ -1:55$ & $ -1:38$ \\ 
2002 Jun 15 &  1 & $ -1:14$ & \nodata \\ 
2002 Jun 16 &  1 & $ -2:41$ & \nodata \\ 
2002 Oct 26 &  1 & $ +2:20$ & \nodata \\ 
2003 Jun 23 &  3 & $ -1:29$ & $ -1:12$ \\ 
\cutinhead{UY Pup:}
2002 Feb 20 &  2 & $ +2:10$ & $ +2:22$ \\ 
2002 Feb 21 & 11 & $ -2:26$ & $ +3:05$ \\ 
2002 Feb 22 &  7 & $ -2:07$ & $ +2:16$ \\ 
2002 Oct 29 &  1 & $ -1:32$ & \nodata \\ 
2002 Oct 30 &  1 & $ -0:09$ & \nodata \\ 
2002 Oct 31 &  1 & $ +0:02$ & \nodata \\ 
2002 Nov 1 &  2 & $ -1:14$ & $ -1:01$ \\ 
2002 Dec 12 &  4 & $ -3:51$ & $ +2:33$ \\ 
2002 Dec 13 &  6 & $ -3:30$ & $ +2:38$ \\ 
2002 Dec 14 &  1 & $ -2:53$ & \nodata \\ 
2002 Dec 16 &  3 & $ -2:20$ & $ +1:07$ \\ 
2002 Dec 19 &  5 & $ +0:19$ & $ +3:27$ \\ 
2003 Feb 1 &  1 & $ -0:43$ & \nodata \\ 
2003 Feb 2 &  1 & $ +0:33$ & \nodata \\ 
\cutinhead{RY Ser:}
2003 Jun 20 &  4 & $ +0:09$ & $ +3:42$ \\ 
2003 Jun 21 &  4 & $ -2:44$ & $ +3:14$ \\ 
2003 Jun 22 &  3 & $ -2:48$ & $ +3:00$ \\ 
2003 Jun 23 &  2 & $ -2:18$ & $ -2:09$ \\ 
2003 Jun 25 &  1 & $ -0:35$ & \nodata \\ 
\cutinhead{CH UMa:}
2000 Jan 10 &  1 & $ +3:22$ & \nodata \\ 
2002 Feb 16 &  1 & $ -0:29$ & \nodata \\ 
2002 Feb 17 &  2 & $ +1:51$ & $ +2:04$ \\ 
2002 Feb 20 &  1 & $ +4:43$ & \nodata \\ 
2002 Feb 21 &  2 & $ -3:20$ & $ +3:10$ \\ 
2002 Feb 22 &  5 & $ -3:24$ & $ +5:13$ \\ 
2002 Jun 14 &  1 & $ +3:34$ & \nodata \\ 
2002 Jun 15 &  1 & $ +3:40$ & \nodata \\ 
2002 Jun 16 &  1 & $ +3:36$ & \nodata \\ 
2002 Jun 17 &  1 & $ +3:45$ & \nodata \\ 
2002 Oct 26 &  1 & $ -2:21$ & \nodata \\ 
2002 Oct 29 &  1 & $ -2:18$ & \nodata \\ 
2002 Dec 13 &  1 & $ +0:32$ & \nodata \\ 
2002 Dec 19 &  4 & $ -1:27$ & $ +1:27$ \\ 
2003 Jan 31 &  1 & $ +3:57$ & \nodata \\ 
2003 Jun 20 &  2 & $ +4:29$ & $ +5:06$ \\ 
2003 Jun 21 &  1 & $ +4:00$ & \nodata \\ 
2003 Jun 23 &  1 & $ +3:56$ & \nodata \\ 
\cutinhead{SDSS 0813+45:}
2002 Jan 19 & 11 & $ -4:15$ & $ +5:03$ \\ 
2002 Jan 20 &  5 & $ -1:12$ & $ +3:39$ \\ 
2002 Jan 21 &  5 & $ -2:42$ & $ +2:49$ \\ 
2002 Jan 22 &  3 & $ -0:29$ & $ +3:39$ \\ 
2002 Jan 24 &  1 & $ +1:38$ & \nodata \\ 
2002 Feb 19 &  1 & $ +0:46$ & \nodata \\ 
\enddata
% \tablenotetext{a}{Emission equivalent widths are counted as positive.}
% \tablenotetext{b}{Absolute line fluxes are uncertain by a factor of about
% 2, but relative fluxes of strong lines
% are estimated accurate to $\sim 10$ per cent.}
% \tablenotetext{c}{From Gaussian fits.}
\end{deluxetable}

\clearpage

\begin{deluxetable}{lrcc}
\tablewidth{0pt}
\tabletypesize{\footnotesize}
\tablecolumns{4}
\tablecaption{Emission Features}
\tablehead{
\colhead{Feature} &
\colhead{E.W.\tablenotemark{a}} &
\colhead{Flux\tablenotemark{b}}  &
\colhead{FWHM \tablenotemark{c}} \\
 &
\colhead{(\AA )} &
\colhead{(10$^{-16}$ erg cm$^{-2}$ s$^{1}$)} &
\colhead{(\AA)} \\
}
\startdata
\cutinhead{TT Crt:} 
H$\gamma$ & $ 14$ & $176$ & 32 \\ 
H$\beta$ & $ 13$ & $165$ & 28 \\ 
NaD & $ -3$ & $-47$ & 11 \\ 
H$\alpha$ & $ 17$ & $237$ & 29 \\ 
\cutinhead{EZ Del:}
           H$\gamma$ & $ 18$ & $ 87$ & 16 \\
  HeI $\lambda 4471$ & $  2$ & $ 11$ &  6 \\
            H$\beta$ & $ 22$ & $ 82$ & 14 \\
  HeI $\lambda 4921$ & $  3$ & $  9$ & 17 \\
  HeI $\lambda 5015$ & $  3$ & $ 10$ & 16 \\
   Fe $\lambda 5169$ & $  1$ & $  5$ & 11 \\
  HeI $\lambda 5876$ & $  5$ & $ 14$ & 13 \\
                 NaD & $ -1$ & $ -2$ &  9 \\
           H$\alpha$ & $ 35$ & $ 85$ & 16 \\
  HeI $\lambda 6678$ & $  3$ & $  8$ & 16 \\
  HeI $\lambda 7067$ & $  2$ & $  4$ & 19 \\
\cutinhead{LL Lyr:}
           H$\gamma$ & $ 38$ & $137$ & 21 \\ 
  HeI $\lambda 4471$ & $ 10$ & $ 33$ & 35 \\ 
            H$\beta$ & $ 58$ & $147$ & 26 \\ 
  HeI $\lambda 4921$ & $  9$ & $ 23$ & 27 \\ 
  HeI $\lambda 5015$ & $ 12$ & $ 30$ & 42 \\ 
   Fe $\lambda 5169$ & $  5$ & $ 14$ & 21 \\ 
  HeI $\lambda 5876$ & $ 16$ & $ 37$ & 29 \\ 
           H$\alpha$ & $ 68$ & $172$ & 24 \\ 
  HeI $\lambda 6678$ & $  8$ & $ 18$ & 31 \\ 
\cutinhead{UY Pup:}
           H$\gamma$ & $  8$ & $191$ & 12 \\ 
  HeI $\lambda 4471$ & $  1$ & $ 35$ & 11 \\ 
            H$\beta$ & $ 10$ & $249$ & 12 \\ 
  HeI $\lambda 4921$ & $  1$ & $ 33$ & 14 \\ 
  HeI $\lambda 5015$ & $  1$ & $ 30$ & 12 \\ 
  HeI $\lambda 5876$ & $  3$ & $ 66$ & 12 \\ 
                 NaD & $ -1$ & $-25$ & 10 \\ 
           H$\alpha$ & $ 17$ & $387$ & 14 \\ 
  HeI $\lambda 6678$ & $  2$ & $ 47$ & 16 \\ 
  HeI $\lambda 7067$ & $  2$ & $ 34$ & 17 \\ 
\cutinhead{RY Ser:}
           H$\gamma$ & $ 26$ & $198$ & 20 \\ 
  HeI $\lambda 4471$ & $  6$ & $ 53$ & 11 \\ 
 HeII $\lambda 4686$ & $  3$ & $ 23$ & 16 \\ 
            H$\beta$ & $ 19$ & $179$ & 12 \\ 
  HeI $\lambda 4921$ & $  1$ & $ 11$ & 11 \\ 
  HeI $\lambda 5015$ & $  2$ & $ 16$ & 12 \\ 
  HeI $\lambda 5876$ & $  5$ & $ 58$ & 11 \\ 
                 NaD & $ -2$ & $-25$ &  9 \\ 
           H$\alpha$ & $ 22$ & $296$ & 12 \\ 
  HeI $\lambda 6678$ & $  3$ & $ 36$ & 17 \\ 
  HeI $\lambda 7067$ & $  2$ & $ 31$ & 20 \\ 
\cutinhead{CH UMa:}
           H$\gamma$ & $ 31$ & $798$ & 12 \\ 
  HeI $\lambda 4471$ & $ 13$ & $314$ & 14 \\ 
            H$\beta$ & $ 35$ & $980$ & 12 \\ 
  HeI $\lambda 4921$ & $  4$ & $109$ & 12 \\ 
  HeI $\lambda 5015$ & $  4$ & $105$ & 10 \\ 
  HeI $\lambda 5876$ & $  9$ & $311$ & 11 \\ 
                 NaD & $ -3$ & $-87$ & 11 \\ 
           H$\alpha$ & $ 36$ & $1268$ & 11 \\ 
  HeI $\lambda 6678$ & $  5$ & $168$ & 15 \\ 
  HeI $\lambda 7067$ & $  3$ & $113$ & 16 \\ 
\cutinhead{SDSS 0813+45:}
           H$\gamma$ & $ 19$ & $ 35$ & 10 \\ 
  HeI $\lambda 4471$ & $ 13$ & $ 22$ & 19 \\ 
            H$\beta$ & $ 35$ & $ 57$ & 12 \\ 
  HeI $\lambda 4921$ & $  2$ & $  4$ & 10 \\ 
  HeI $\lambda 5015$ & $  3$ & $  4$ &  9 \\ 
  HeI $\lambda 5876$ & $  6$ & $ 11$ & 11 \\ 
                 NaD & $ -3$ & $ -6$ & 10 \\ 
           H$\alpha$ & $ 30$ & $ 55$ & 11 \\ 
  HeI $\lambda 6678$ & $  4$ & $  6$ & 17 \\ 
  HeI $\lambda 7067$ & $  3$ & $  5$ & 17 \\ 
\enddata
\tablenotetext{a}{Emission equivalent widths are counted as positive.}
\tablenotetext{b}{Absolute line fluxes are uncertain by a factor of about
2, but relative fluxes of strong lines
are estimated accurate to $\sim 10$ per cent.}
\tablenotetext{c}{From Gaussian fits.}
\end{deluxetable}

\clearpage
\begin{deluxetable}{lrcrc}
\tablewidth{0pt}
\tablecolumns{5}
\tablecaption{Radial Velocities}
\tablehead{
\colhead{Time\tablenotemark{a}} &
\colhead{$v_{\rm abs}$} &
\colhead{$\sigma$} &
\colhead{$v_{\rm emn}$} &
\colhead{$\sigma$} \\
\colhead{} &
\colhead{(km s$^{-1}$)} &
\colhead{(km s$^{-1}$)} &
\colhead{(km s$^{-1}$)} &
\colhead{(km s$^{-1}$)} 
}
\startdata
\cutinhead{TT Crt:}
51550.0607  & $   36$ &   16 & $  -75$ &   21 \\
51550.0648  & $  -12$ &   17 & $  -71$ &   26 \\
51641.8325  & $   44$ &   15 & $ -118$ &   13 \\
52262.0528  & $ -203$ &   14 & $   74$ &   23 \\
52262.0599  & $ -219$ &   14 & $    4$ &   27 \\
52268.9147  & $  190$ &   19 & $ -177$ &   20 \\
52268.9218  & $  181$ &   18 & $ -188$ &   20 \\
52268.9290  & $  196$ &   17 & $ -210$ &   20 \\
52270.9242  & $ -271$ &   19 & $   -2$ &   27 \\
\enddata
\tablenotetext{a}{Heliocentric Julian data of 
mid-exposure, minus 2 400 000.}
\tablecomments{Emission (H$\alpha$) and absorption radial velocities.  
A sample is shown here; full version 
is available in the electronic version of the paper.}
\end{deluxetable}

\clearpage

\begin{deluxetable}{lllrrcc}
\tablecolumns{7}
% \tabletypesize{\footnotesize}
\tablewidth{0pt}
\tablecaption{Fits to Radial Velocities}
\tablehead{
\colhead{Data set} & 
\colhead{$T_0$\tablenotemark{a}} & 
\colhead{$P$} &
\colhead{$K$} & 
\colhead{$\gamma$} & 
\colhead{$N$} &
\colhead{$\sigma$\tablenotemark{b}}  \\ 
\colhead{} & 
\colhead{} &
\colhead{(d)} & 
\colhead{(km s$^{-1}$)} &
\colhead{(km s$^{-1}$)} & 
\colhead{} &
\colhead{(km s$^{-1}$)} \\
}
\startdata
TT Crt (abs) & 297.0254(9) & 0.2683522(5) &  212(5) & $-23(3)$ & 36 &  15 \\ 
TT Crt (emn) & 297.163(3) & \nodata  &  124(8) & $-77(6)$ & 36 &  27 \\ 
% OK, basically the one selected from Shafter.
\\[0.5ex]
EZ Del (emn)\tablenotemark{c} & 442.917(2) & 0.2234(5) &  91(7) & $-6(5)$ & 48 &  21 \\ 
% no shafter 
\\[0.5ex]
LL Lyr (emn)\tablenotemark{c} & 042.079(3) & 0.249069(4) &  75(5) & $-35(4)$ & 97 &  17 \\ 
% no shafter 
\\[0.5ex]
UY Pup (abs) & 577.183(3) & 0.479269(7) &  102(4) & $ 45(3)$ & 46 &  14 \\
UY Pup (emn) & 576.948(5) & \nodata &  96(5) & $ 28(4)$ & 49 &  18 \\ 
% this is the one used for shafter vels.
\\[0.5ex]
RY Ser (abs) & 812.788(3) & 0.3009(4) &  87(6) & $-10(4)$ & 14 &  12 \\ 
RY Ser (emn) & 812.639(6) & \nodata   &  73(10) & $-30(6)$ & 14 &  20 \\ 
% This is the dynamical velocity set.
\\[0.5ex]
CH UMa (abs) & 442.787(3) &  0.3431843(6)  &  76(3) & $-15(2)$ & 28 &   8 \\ 
CH UMa (emn)\tablenotemark{c} & 442.616(9) & \nodata &  33(5) & $-14(4)$ & 28 &  13 \\ 
% This is the dgau5 Halpha, b/c S/N is so bad with the gau2.  Gau2
% suggests a larger K by somewhat.  Phase of T0(emn) is 0.502(8).
\\[0.5ex]
SDSS0813 (abs) & 296.080(5) & 0.2890(4)\tablenotemark{d}  &  54(7) & $-29(4)$ & 23 &  15 \\ 
SDSS0813 (emn)\tablenotemark{c} & 295.942(6) & \nodata &  32(5) & $-22(3)$ & 26 &  11 \\ 
\enddata
\tablecomments{Parameters of least-squares sinusoid fits to the radial
velocities, of the form $v(t) = \gamma + K \sin(2 \pi(t - T_0)/P$.
Where both emission and absorption velocities are available, the 
period quoted is the weighted average of the periods derived from 
separate fits to the two data sets, and the period is only given
on the first line.}
\tablenotetext{a}{Heliocentric Julian Date minus 2452000.  The epoch is chosen
to be near the center of the time interval covered by the data, and
within one cycle of an actual observation.}
\tablenotetext{b}{Root-mean-square residual of the fit.}
\tablenotetext{c}{In these cases the emission line velocities which were
fitted were not derived using the double-gaussian convolution; in 
EZ Del and LL Lyr the lack of secondary-star velocities precluded
checking the phase, and in RY Ser and CH UMa the lines were narrow
enough that the double-gaussian method did not offer any advantage.}
\tablenotetext{d}{There is slight ambiguity in the adopted period, see
the discussion in the text.}
\end{deluxetable}

\clearpage

\begin{deluxetable}{llrrrrcr}
\tabletypesize{\scriptsize}
\tablewidth{0pt}
\tablecolumns{8}
\tablecaption{Inferences from Secondary Stars}
\tablehead{
\colhead{Star} &
\colhead{Type} &
\colhead{Synthetic $V$}  &
\colhead{Assumed $M_2$\tablenotemark{a}} &
\colhead{Deduced $R_2$} &
\colhead{$M_V$\tablenotemark{b}} &
\colhead{$A_V$} &
\colhead{Distance} \\ 
 &
 & 
\colhead{(mag)} &
\colhead{$M_{\odot}$} &
\colhead{$R_{\odot}$} &
\colhead{(mag)} &
\colhead{(mag)} &
\colhead{(pc)} \\
}
\startdata
TT Crt & K5 $\pm$ 1 & $16.8 \pm 0.3$ & $0.65 \pm 0.15$ & $0.69 \pm 0.06$
& $7.3 \pm 0.4$ & 0.1 & $760 (+200,-160)$ \\
EZ Del & M1.5 $\pm$ 0.5 & $20.0 \pm 0.4$ & $0.45 \pm 0.10$ & $0.52 \pm
0.07$ & $9.5 \pm 0.5$ & $0.5 \pm 0.2$ & 1000($+380,-280)$\\ 
LL Lyr & M2.5 $\pm$ 1.5 &  $20.0 \pm 0.4$\tablenotemark{c} & $0.39 \pm
0.14$ & $0.55 \pm 0.07$ & $9.9$\tablenotemark{c} & 0.2 & 960($+420,-300$) \\
UY Pup & K4 $\pm$ 2 & $17.3 \pm 0.5$ & $0.65 \pm 0.25$ & $1.0 \pm 0.2$ & $6.3 \pm 0.8$ &
0.5 & 1300 ($+600,-500$) \\
RY Ser & K5 $\pm$ 1 & $17.4 \pm 0.3$ & $0.6 \pm 0.1$ & $0.74 \pm 0.05$ & 
$7.2 \pm 0.4$ & $1.2 \pm 0.3$ & 620($+240,-170$) \\
CH UMa & K5.5 $\pm$ 1 & $15.9 \pm 0.4$ & $0.6 \pm 0.15$ & $0.78 \pm 0.08$ & 
$7.3 \pm 0.6$ & 0.18 & 480($+180, -130$) \\
SDSS0813+45 & K5.5 $\pm$ 1 & $19.1 \pm 0.4$  & $0.63 \pm 0.10$ & 
$0.72 \pm 0.04$ & $7.4 \pm 0.3$ & 0.15 & $2100 \pm 500$ \\
\enddata
\tablenotetext{a} {Note carefully that these masses are not measured,
but are estimates guided by the models of \citet{bk00}.  They are 
used {\it only} to constrain $R_2$, which depends only on the cube
root of $M_2$, so this does not contribute substantially to the error
budget.}
\tablenotetext{b} {Absolute visual magnitude inferred for the secondary
alone, on the basis of surface brightness and Roche lobe size (see
text).}
\tablenotetext{c} {The error estimate for LL Lyr is complicated because
the secondary flux is correlated with the spectral type.  Fortunately,
these errors tend to compensate in the distance calculation, because a 
later-type star, with stronger
features, is inferred to contribute less to the summed spectrum, but
is also inferred to be intrinsically fainter.  The final distance
uncertainties include this effect.}
\end{deluxetable}

\clearpage

\begin{figure}
% \epsscale{0.90}
\plotone{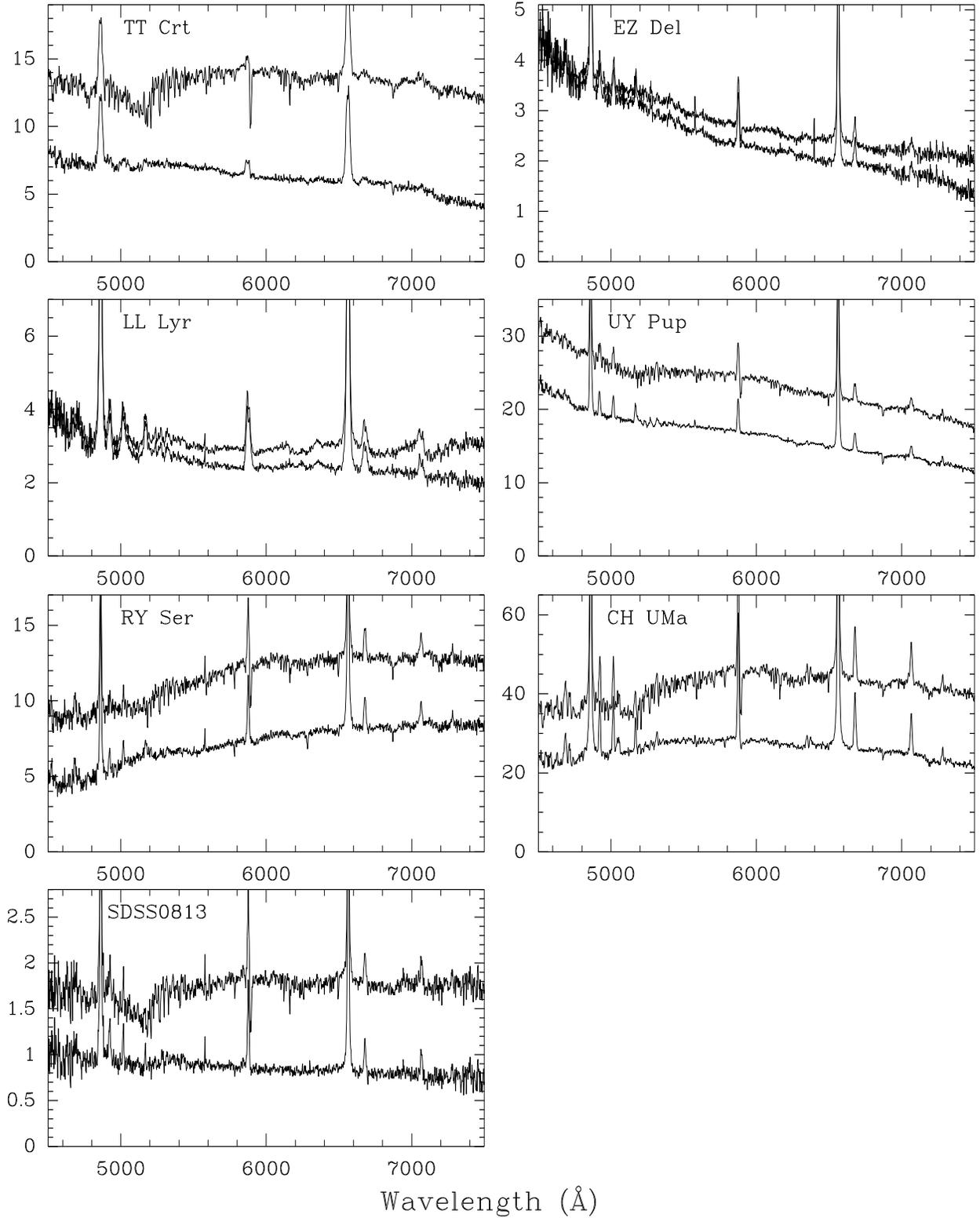}
\caption{A montage of spectra.  The vertical scale in each plot is in units
of $10^{-16}$ erg s$^{-1}$ cm$^{-2}$ \AA $^{-1}$, subject to 
calibration uncertainties of some tens of percent.  The 
lower trace in each panel shows the data after a scaled
late-type star has been subtracted away (see text and Table 5).
In all cases except EZ Del and LL Lyr the original spectra
were shifted into the rest frame of the secondary star before
averaging.
}
\end{figure}

\clearpage

\begin{figure}
\epsscale{1.0}
\plotone{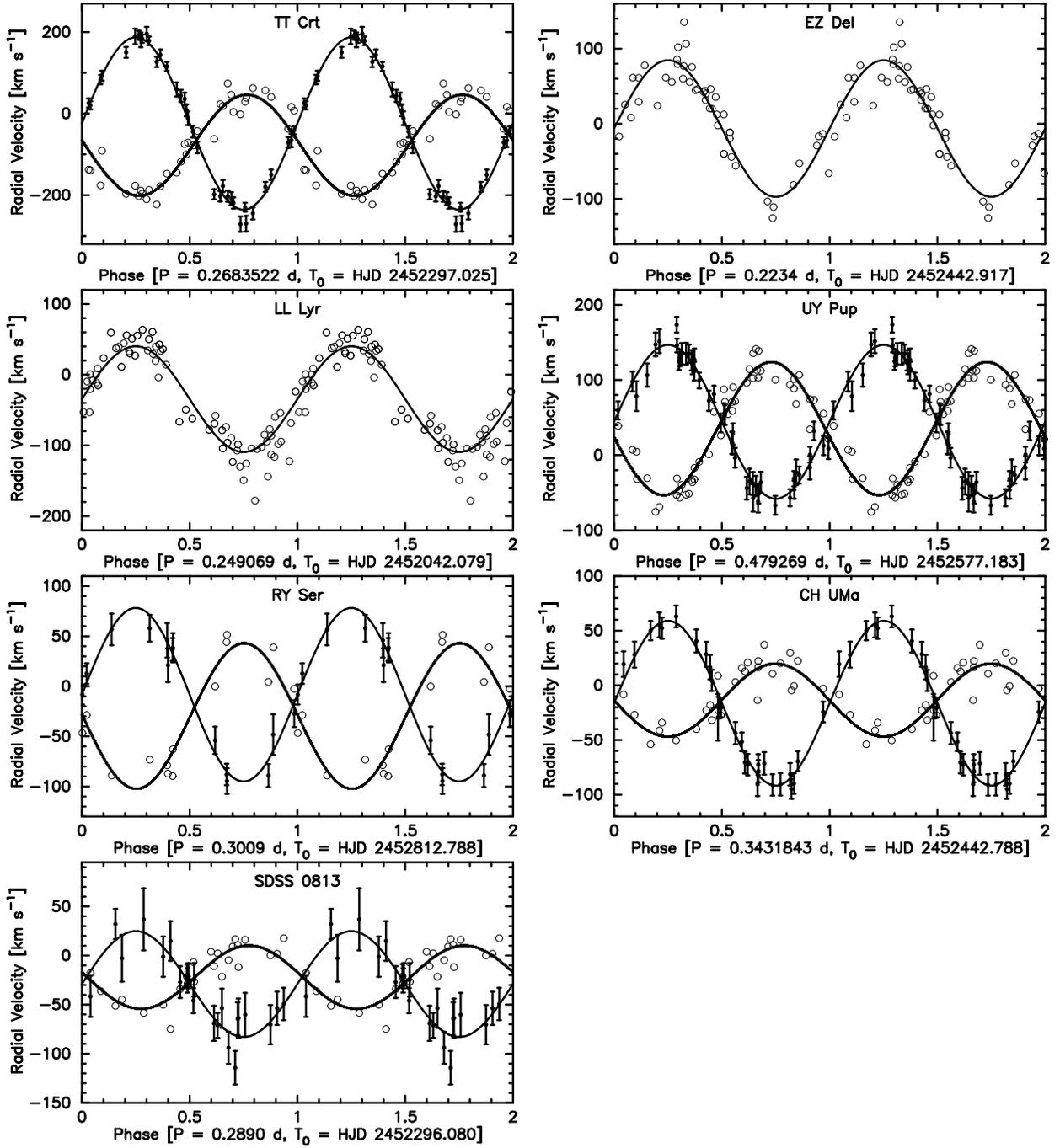}
\caption{Absorption (solid dots with error bars) and emission (open
circles) radial velocities folded on the adopted orbital periods.
Best-fit sinusoids are superposed.  All data are shown twice
for continuity.
}
\end{figure}

\end{document}